**Functional Amyloid Fibrils as Versatile Tools for Novel Biomaterials**

Shayan Mortazavi[a], Mehrnoosh Neshatian[b], Laurent Bozec [b], Hadis Zarrin [c] Mahshid Kalani [*c]

[a] Department of Science, York University, Canada.
[b] Department of Dentistry, University of Toronto, Canada.
[b] Department of Chemical Engineering, Toronto Metropolitan University, Canada.

**Abstract:**

Functional amyloid fibrils, once primarily associated with amyloidosis, are now recognized for their exceptional potential as biomaterials due to their unique structural features, including remarkable mechanical strength, high stability, and self-assembly capabilities. This review highlights their transformative applications across a wide range of industries, from cutting-edge drug delivery systems and next-generation biosensors to tissue engineering, surface technologies, energy storage, and environmental solutions. Their versatility extends into innovative sectors like information transfer systems, cell adhesion, protein fusion, food technology, and novel catalytic systems. Despite significant progress, critical gaps remain in the research. This review not only consolidates current applications but also underscores the vast potential for future advancements, positioning functional amyloids as key players in emerging biomaterials technologies.

**Introduction**

Amyloid fibrils, once primarily associated with neurodegenerative diseases, have gained interest for their unique structural characteristics and potential biological roles. Traditionally viewed as pathological, these *β*-sheet-rich protein aggregates [1–5] are now recognized for their functionality across various biological systems, including biofilm formation in bacteria and hormonal regulation in higher organisms. [6–12]

Recent research highlights the transition from a pathological to a functional perspective of amyloid fibrils, paving the way for innovative applications in biotechnology and materials science. [13] By leveraging the self-assembly properties and biocompatibility of functional amyloids, advancements in drug delivery, tissue engineering, and nanotechnology are emerging. Ryadnov *et al*. [14] explored the properties of amyloid fibrils derived from pentapeptide-based peptide amphiphiles, further underscoring their versatility in various fields.

Functional amyloids exhibit unique structural features, including fibrillar formations with a central cross-β scaffold. [15] This structure facilitates novel biological functions and demonstrates how the supramolecular framework can accommodate diverse amino acid sequences while regulating assembly selectivity. For instance, an amphipathic *β*-strand can bind tightly to lipids when part of an amyloid fibril, thanks to cooperative effects and avidity resulting from repetitive binding pockets on the amyloid surface. [16]

Functional amyloid proteins can transition directly from soluble monomers to insoluble amyloid states, bypassing intermediates like protofibrils, and are characterized by a prion-like replication mechanism. [15,17] Under normal physiological conditions, a protein's native conformation is dictated by the system's lowest Gibbs free energy. [18,19] However, functional amyloids can stabilize at high concentrations due to dense intermolecular hydrogen bonding and steric interactions. [20]

The remarkable mechanical properties of amyloid fibrils, including a high Young's modulus, [21–24] make them attractive for material applications. They function as protective coatings and catalytic materials and possess an exceptionally high aspect ratio, [25,26] which enables fragmentation and proliferation through fission, aiding epigenetic information transfer. [27,28]

Functional amyloids are vital across different biological systems. In bacteria, they enhance biofilm formation, [29] while in fungi, they facilitate cellular adhesion [30] and signal transduction. [31] In mammals, functional amyloids are crucial for hormone storage and release [32] and play roles in innate immunity [33] by trapping pathogens. [34] Their prion-like propagation supports phenotypic inheritance, offering a mechanism for epigenetic regulation. [29]

Exploring functional amyloids not only expands our understanding of protein chemistry and cellular biology but also opens new avenues for developing innovative biomaterials and therapeutic strategies. Amyloids' unique properties can be harnessed toward future applications in nanotechnology, drug delivery, and synthetic biology.

1. **Structural Properties of Functional Amyloids**

Functional amyloid fibrils possess unique structural features that set them apart from pathological amyloids, contributing to their varied functionalities. Key characteristics include stability, polymorphism, and assembly dynamics are summarized in

**Table *1***.

**Table 1: Key Features of Functional Amyloid Fibrils**

| Key Characteristics | Description | References |
|---|---|---|
| Stability | Functional amyloid fibrils are exceptionally stable, enabling them to withstand environmental stresses like pH changes, temperature fluctuations, and proteolytic degradation. Their extensive cross-$\beta$-sheet structure provides mechanical strength and durability. | [35–41] |
| Polymorphism | They exhibit polymorphism, allowing various structural conformations while preserving functionality. This arises from differences in $\beta$-sheet packing, enabling interactions with diverse molecular partners. | [5,7,12,35,42] |
| Assembly Dynamics | The assembly process involves nucleation, elongation, and maturation, influenced by factors like protein concentration, pH, ionic strength, and cofactors. This regulated assembly allows precise control over fibril formation. | [12,39,40,42,43] |

The interplay of stability, polymorphism, and dynamic assembly allows functional amyloid fibrils to serve versatility and functionality as resilient scaffolds in biological processes like cell adhesion, biofilm formation, host defense, [5,12] and molecular recognition. These attributes make functional amyloids valuable in biotechnology and biomedicine, leading to innovations in biomaterials, drug delivery systems, and tissue engineering scaffolds.

## 2. Functional amyloid fibrils Application

Functional amyloid fibrils have recently emerged as versatile biomaterials with a myriad of industrial applications, captivating the attention of researchers and industries alike. Their unique structural properties, including high stability and mechanical strength, make them attractive candidates for diverse applications:

### 2.1 Harnessing Functional Amyloid Fibrils for Surface Engineering Applications

Amyloid fibrils, known for their highly ordered and stable $β$-sheet structures, have been increasingly explored for their potential in facilitating monolayer formation at surfaces, similar to the function of hydrophobins. These fibrils can self-assemble [44] into robust monolayers at various interfaces, creating highly uniform coatings that [45–48] significantly alter surface properties. By leveraging the intrinsic stability and surface activity of amyloid fibrils, researchers have been able to develop coatings that enhance adhesion, [45,49–52] modify wettability, [46,50–52] and provide protective barriers on diverse substrates. [46,49–51] For example, amyloid fibril monolayers can transform hydrophobic surfaces to hydrophilic ones, [53–56] akin to hydrophobin behavior, [53–55] enabling applications in biotechnology, materials science, and medicine. The versatility of amyloid fibrils in forming functional monolayers opens up new avenues for designing advanced materials with tailored surface characteristics, from anti-fouling coatings to biosensors and drug delivery systems. [57–60] Similarly, numerous fungi generate hydrophobins, amphipathic proteins that can form $β$-sheet rich fibrils at air-water interfaces. [61] These fibrils are believed to serve a protective function in fungal structures like spores [61,62] and fruiting bodies, [61,63] prevent desiccation and environmental damage. Hydrophobins facilitate the attachment of fungi to various surfaces, [61,64] ,aiding in colonization and growth, [65–67] and create hydrophobic surfaces, [68–71] which can be beneficial in environments with varying moisture levels. Thus, both amyloid fibrils and hydrophobins exemplify how $β$-sheet rich fibril formation at interfaces can be harnessed to create functional monolayers with diverse applications.

## 2.2 Unveiling Nature's Architect: Functional Amyloid Fibrils in Streptomycetes' Aerial Hyphae Formation (Exemplified by in chaplins)

Amyloid fibrils, exemplified by proteins like Chaplins in Streptomycetes, [72,73] are essential for the formation of aerial hyphae, [72–74] pivotal structures for the reproductive and survival strategies of these filamentous bacteria. The self-assembly of Chaplin amyloid fibrils at air-water interfaces [75–78] provides a scaffold that promotes the emergence and extension of aerial hyphae from the substrate. By forming a hydrophobic layer, amyloid fibrils facilitate the cohesion [75–78] and stability [76–78] of aerial hyphae, allowing them to extend and protrude into the air, ultimately leading to the formation of reproductive structures such as spores. This process is essential for *Streptomycetes*' adaptation to changing environmental conditions and their dispersal in diverse ecological niches. The involvement of amyloid fibrils in aerial hyphae formation highlights their functional significance beyond their conventional roles, emphasizing their importance in microbial physiology and ecological processes.

## 2.3 Harnessing the Potential of Amyloid Fibrils as Innovative Information Carriers

Amyloid fibrils present exciting opportunities in information careers for data storage and processing.[79–83] Their ordered structures make them promising candidates for novel storage devices, [84] enabling high-density [80,81,85,86] and long-term solutions. [80,81] Additionally, they may have applications in bioinformatics, aiding the organization and transmission of biological information within cells. [35,87–89] As research advances, amyloid fibrils could revolutionize data storage and processing, benefiting information technology and biotechnology. Furthermore, they play roles in biological processes, such as the fungal immune system [27,90,91] and long-term memory formation. [27,92] Yeast prions may contribute to heritable information transfer, [28,93,94] while amyloid fibrils are crucial for melanosome maturation, [95,96] highlighting their diverse biological functions.

## 2.4 Pioneering Reservoirs and Storage Units in Biotechnology

Functional amyloid fibrils efficiently store compounds, particularly in biological systems like bacterial biofilms. Their repetitive binding sites create a structured environment, allowing controlled storage of molecules that regulate vital biological processes. [97] In bacterial biofilms, for instance, amyloid fibrils sequester quorum-sensing molecules, [98–100] coordinating bacterial activity. Similarly, in certain fungi, [27,101] like yeast, HET-s prion amyloids participate in immune responses and long-term memory storage, transmitting information [27,101] through stable aggregated states.

The neuronal variant of CPEB (cytoplasmic polyadenylation element-binding protein) [27] forms amyloid-like aggregates that maintain long-term synaptic changes essential for memory, suggesting amyloids may act as biochemical substrates for memory storage and transmission. [101]

Peptide hormone amyloids also play a significant role in the formation of secretory granules in endocrine cells, [102] where they store hormones in a condensed, inactive form. [103,104] These amyloids regulate the controlled release of hormones. [25] Riek *et al.* [32,105] demonstrated that protein hormones form amyloids within secretory granules, effectively releasing their contents during secretion. The aggregation of *provasopressin* [106] into amyloids is essential for its storage and release, indicating a broader regulatory role of amyloids in endocrine function. [3,113]

However, amyloid formation is not always beneficial. Studies have shown that mutations leading to the aggregation of *provasopressin* [107] can result in diabetes-related amyloidosis. Additionally, hormone amyloids, such as calcitonin and insulin, are linked to excessive hormone production in certain cancers, including endocrine tumors. [108]

## 2.5 Harnessing Amyloid Fibrils as Biofilm Matrix scaffold

Amyloid fibrils play a critical role in the biofilm matrix scaffold, providing structural integrity and stability to microbial communities. [29,109–111] These fibrils form an interwoven network that enhances the mechanical strength and resilience of the biofilm, allowing it to withstand various environmental stresses. Amyloid fibers serve as a robust platform facilitating interactions among neighboring cells and surfaces. [112,113] They play a pivotal role in biofilm formation and cell adhesion by imparting rigidity and stiffness to the matrix. [114] Within the biofilm matrix, amyloids aid in adhering to both abiotic and biotic surfaces, [115] elevate hydrophobicity [116–118], and facilitate colonization. [119–123] Additionally, they bolster structural integrity, [99,109,124–126] confer resistance to environmental pressures, [36,99,121,127,128] shield against phage particles, [99,129–132] and defend against enzymes that degrade the matrix. [99,127,133]

**Table** *2* summarized the different studied of the functional amyloid fibril's application in biofilm formation.

**Table 2: Functional Amyloid Fibrils in Biofilm Formation**

| Fibril System | Key Findings | References |
|---|---|---|
| β-lactoglobulin amyloid fibrils | these fibrils contribute to the formation of robust, stable films, making them attractive for use in developing new biomaterials or coatings. | [134] |
| phenol-soluble modulins (PSMs) of Staphylococcus aureus | PSM fibrils are crucial for resilient biofilm formation, enhancing bacterial colonization and resistance to environmental stresses. They play a key role in pathogenicity and persistence in biofilm-related infections. Mutants lacking PSMs are more susceptible to breakdown by enzymes and mechanical stress, highlighting their importance in biofilm stability. | [135] |
| phenol-soluble modulins (PSMs) of Staphylococcus aureus | There is dual functionality of soluble PSMs and polymerized PSMs in biofilm dynamics. Soluble PSMs can disperse biofilms while polymerized PSMs form β-sheet-rich aggregates that provide structural support. This points to the complex interplay between PSM structure and environmental conditions affecting biofilm formation. | [136–140] |
| Bap (biofilm-associated protein) of Staphylococcus aureus | Bap forms insoluble amyloid-like aggregates at low $Ca^{2+}$ levels and acidic pH, aiding biofilm formation, while higher $Ca^{2+}$ concentrations stabilize its conformation, inhibiting amyloid aggregation and enhancing biofilm resilience. | [141,142] |
| Curli fiber of Escherichia coli | Enhanced Curli fiber production in Escherichia coli improves biofilm formation, strengthens surface attachment, increases hydrophobicity, and promotes pathogenicity by facilitating attachment to host proteins and surfaces. | [143,144], [143,145] |
| Curli fiber of Escherichia coli | Curli fibers is applied in creating a biocatalytic biofilm that immobilizes the enzyme α-amylase, enhancing its catalytic efficiency for various applications. | [146] |
| Extracellular fibers of Pseudomonas | They are similar to curli fibrils and are encoded by the fapABCDEF operon, indicating a conserved mechanism for biofilm formation across different bacterial species | [109] |
| TasA of Bacillus subtilis and Bacillus cereus | These fibers are crucial for the extracellular matrix of biofilms, promoting cell cohesion and providing structural support to maintain biofilm integrity. | [147,148] |
| Salmonella curli fibers | Salmonella curli fibers play a critical role in biofilm stability by strongly binding to extracellular DNA, protecting it from degradation by DNases. Nucleic acids facilitate the polymerization of curli fibers, accelerating the formation of stable complexes from CsgA monomers. This mechanism enhances the stability of Salmonella enterica serovar Typhimurium biofilms, demonstrating how DNA molecules contribute to the formation of robust bacterial biofilm structures. | [149,150] |
| Mycobacterium tuberculosis | Pili involved in biofilm lifestyle, potential target for biofilm inhibition in tuberculosis. | [151,152] |

| | |
|---|---|
| Fap fibrils in Pseudomonas aeruginosa | Fap fibrils contribute to biofilm structure, enhancing adhesion to surfaces and providing protection against environmental stresses, facilitating persistent infections. |

### 2.6 Amyloid Fibrils as a Novel Tools for Cell adhesion

Amyloid fibrils are crucial for cell adhesion, providing a versatile scaffold that facilitates strong interactions between cells and surfaces. Their stable $\beta$-sheet structures enable the formation of robust biofilms, with curli fibers in *E. coli* and phenol-soluble modulins (PSMs) in *S. aureus* playing key roles in initial biofilm formation, [136,143] allowing bacteria to adhere firmly to surfaces and withstand shear forces. Beyond microbial systems, synthetic amyloid fibrils mimic the extracellular matrix, promoting adhesion of mammalian cells, [50,154] which is beneficial in tissue engineering and regenerative medicine. Their biocompatibility [50,155] and customizable surface properties make them suitable for medical implants, enhancing integration with biological tissues and reducing rejection risks.

In addition, the adhesin P1 from *Streptococcus mutans*, a cariogenic pathogen, exhibits amyloidogenic properties [156,157] with an extended fibrillar arrangement that includes a type II helix and globular domains, [158,159] forming characteristic $\beta$-sheet-rich fibers. [160–163] This protein is found covalently attached to peptidoglycan [164,165] and in the culture medium, [166,167] has a crystal structure that reveals an extended fibrillar arrangement. [168,169]

Similarly, Als proteins in yeast, like *Saccharomyces cerevisiae* and *Candida albicans*, [170,171] leading to form nanodomains [172,173] or adhesin clusters [172,174] through amyloid-like interactions, essential for biofilm formation. Research indicates that mutations in amyloid-forming regions of Als proteins can hinder cellular aggregation and biofilm development, [172,175] highlighting potential therapeutic intervention targets in pathogenic biofilm formation. [176–181]

### 2.7 Exploring Amyloid Fibrils as Promising Therapeutic Agents for Intestinal Inflammatory Disorders

Recent research has highlighted the significant role of amyloid fibrils in exacerbating inflammatory responses in the gut, [182] suggesting they are a promising target for therapeutic intervention in intestinal inflammatory disorders such as inflammatory bowel disease (IBD). [183] Various researchers have investigated amyloid fibril inhibitors, including small molecules, peptides, and natural compounds, demonstrating their potential in reducing inflammation and restoring gut homeostasis. [184] For instance, targeting amyloid fibrils with specific inhibitors has shown efficacy in mitigating inflammation and improving clinical outcomes in preclinical models of Crohn's disease and ulcerative colitis. [185] By disrupting amyloid fibril formation, these therapeutic agents can lower inflammatory markers and combat gut inflammation, [186,187] paving the way for novel treatments for chronic intestinal conditions. Also, it is found that curli fibrils derived from both pathogenic and commensal microbiota suggested bacterial amyloids as activate TLR2, leading to the production of immunomodulatory cytokines that alleviate inflammation in a colitis mouse model. [188] However, further research is required to fully explore the immunomodulatory role of bacterial amyloids and assess their potential as therapeutic options for chronic inflammatory conditions.

This emerging approach underscores the therapeutic potential of amyloid fibril inhibitors in managing and potentially alleviating the burden of intestinal inflammatory disorders.

### 2.8 Amyloid Fibrils as Molecular Templates: Unlocking Their Potential in Small Chemical Molecule Assembly

Amyloid fibrils can serve as scaffolds [189] for the self-assembly of nanostructures, highlighting their potential use as templates [190] for the organization of small molecules due to their highly ordered and repetitive structures.

Functional amyloid fibrils have garnered interest for their ability to act as templates for organizing small chemical molecules, leveraging their highly ordered and repetitive structures. These fibrils exhibit a robust, $\beta$-sheet-rich architecture, which allows for the precise arrangement of molecules along their surfaces. This property makes them ideal scaffolds for templating the assembly of nanostructures and organizing small chemical entities—for instance, Bauer et al.[190] demonstrated that amyloid fibrils could guide the self-assembly of hexagonal protein lattices at interfaces, showcasing their potential in molecular templating applications. Similarly, Lu et al. [189] highlighted the utility of amyloid fibrils in nanotube self-assembly, where the fibrils served as a backbone for the organized attachment of chemical species, facilitating the creation of complex nanostructures.

These studies underscore the versatility of amyloid fibrils in nanotechnology and materials science, where their ability to act as precise templates can be harnessed for innovative applications in drug delivery, biosensing, and the fabrication of novel biomaterials.

### 2.9 Revolutionizing Biosensors and Diagnostics

Amyloid fibrils can be employed in biosensors for detecting specific biomolecules due to their structural stability and ability to bind to target molecules, making them ideal for diagnostic applications. [161–163,170] They can be engineered with specific binding sites, enhancing sensitivity and specificity in detecting disease biomarkers [50,191–193] and environmental pollutants. [162,170–172,175] Their self-assembling nature allows for the creation of stable structures on sensor surfaces, improving the reproducibility and performance of biosensors. [45,50,194]

Amyloid fibrils also facilitate signal amplification in biosensors; their aggregation upon binding to target molecules enhances detection sensitivity. [171–175] Their versatility allows them to be tailored for various biosensing applications, which is crucial for specific diagnostic needs. [195–198] Additionally, the dynamic nature of amyloid fibrils enables real-time monitoring of binding events, with structural changes translating into measurable signals. [199–202]

Díaz-Caballero et al. [203] demonstrated the assembly of amyloid-like fibers from prion-inspired heptapeptides using the biotin-streptavidin system under mild conditions. These scaffolds were functionalized with various streptavidin conjugates, including gold nanoparticles, quantum dots, and enzymes like peroxidase and phosphatase. They also modified the amyloid-like nanostructures with glucose oxidase and horseradish peroxidase, establishing a modular approach for developing functionalized amyloid-based nanomaterials, such as a glucose biosensor. The application of Functional amyloid fibrils in biosensors are categorized in **Table 3**.

**Table 3: Functional amyloid fibrils applied in novel biosensors.**

| Type of Biosensor | Description | References |
|---|---|---|
| Fluorescence Biosensors | Amyloid fibrils can be engineered with fluorescent tags to detect biomolecules and pollutants. Binding events alter fluorescence intensity or spectral properties, which can be quantified. | [204–211] |
| Electrochemical Biosensors | Amyloid fibrils immobilized on electrode surfaces enable detection via changes in electrical conductivity upon target binding. | [212–223] |
| Optical Biosensors | Integrated into optical sensing platforms, amyloid fibrils provide label-free detection with high sensitivity. Binding leads to refractive index changes, allowing real-time detection. | [205,224–236] |
| Mechanical Biosensors | Incorporated into nanomechanical devices, amyloid fibrils detect mechanical stress or deformation caused by target binding. | [237–245] |
| Chemical Biosensors | Functionalized amyloid fibrils with specific ligands allow rapid detection of analytes. | 236, 284-289 |

### 2.10 Expanding Horizons: Amyloid Fibrils as Innovations in Drug Delivery

Amyloid fibrils are engineered for drug delivery applications due to their self-assembling nature, enabling effective drug encapsulation and controlled release at targeted sites, potentially improving efficacy and reducing side effects. [246–251] TGF-β3/insulin crystallization establishes a sustained-release mechanism, where micro-crystals act as reservoirs for gradual drug release. [252] Maji *et al.* [253] demonstrated that amyloids can serve as long-acting drug depots, maintaining stability and supporting sustained release of biologically active peptides.

While amyloid-based formulations offer convenience and maintain drug dosage within target ranges, [254,255] challenges include ensuring non-toxicity, [256,257] functional monomer release, [258,259] and minimal interaction with disease-related amyloid proteins. [260–263]

Research on the killer peptide (KP) revealed its ability to self-assemble into fibril-like structures, protected by 1,3-$\beta$-glucans, ensuring controlled release and targeted delivery. [264] Gupta *et al.* [265] developed a variant called Supramolecular Insulin Assembly II (SIA-II), which effectively regulated glucose levels in diabetic models for extended periods.

Self-assembling nanofibrils, $\beta$ such as RADA16 hydrogel, function as slow-release carriers for various proteins. [266,267] The cross-$\beta$-sheet-rich structure's stable core protects drug molecules from heat [268,269] and enzyme degradation, [270,271] highlighting the versatility and efficiency of these nanofibrils in drug delivery. Additionally, α-synuclein's unique fibril morphologies enable nanomatrix applications in enzyme entrapment. [272] A nanoporous *β*(2)-microglobulin matrix, linked to dialysis-related amyloidosis, demonstrates promise in drug delivery and tissue engineering due to its high surface area and selective disintegration. [273]

## 2.11 Amyloid Fibrils as Versatile Platforms for Protein Fusion Applications

Amyloid fibrils are essential in protein fusion applications, enhancing the stability and solubility of fusion proteins. By fusing amyloid-forming peptides with proteins, multifunctional biomolecules are created. [274] These fusion aids in solubilizing aggregation-prone proteins, directing them into insoluble fibrils, thus reducing misfolding and cytotoxicity. [275] The *β*-sheet-rich structure of amyloid fibrils protects fused proteins from enzymatic degradation, [276,277] improving therapeutic efficacy. [278] Amyloid fibril-based protein fusions can be employed to design biomaterials with adjustable properties for tissue engineering and regenerative medicine, aiding in the repair and regeneration of damaged tissues and organs. [279]

Enzyme fusions also exhibit enhanced stability [280,281] and catalytic activity, [282,283] facilitating their immobilization for biocatalytic uses. [284] For example, a fusion of cytochrome b562 with an SH3 dimer formed classical amyloid fibrils and bound metalloporphyrins, demonstrating spectroscopic similarities to wild-type cytochrome. [285]

## 2.12 Harnessing Amyloid Fibrils for Epigenetic Information Transfer

Epigenetic information involves inheritable changes in gene expression without altering the DNA sequence. [286] These changes arise from mechanisms such as DNA methylation methylation, [287] histone modification, [288] and non-coding RNAs. [289] Amyloid fibrils can transmit epigenetic information through fragmenting and proliferating, similar to single-cell organisms. [290] Prion-like proteins act as epigenetic regulators, switching conformational states in response to environmental signals, [291,292] which influences gene expression. [293] Halfmann *et al.* [28] discussed prions in epigenetic inheritance, [28] while Cushman *et al.* focused on prion-like mechanisms in disorders [294]. The neuronal CPEB isoform in Aplysia supports long-term memory through self-propagation. [27,295–297] In yeast, prion proteins like [PSI+] and [URE3] create heritable phenotypic changes. [93,298–300] Fiumara *et al.* explored coiled-coil structures in prions, indicating their role in non-nucleic acid information transfer. [301]

## 2.13 Harnessing Amyloid Fibrils for Food Industries

Amyloid fibrils are gaining traction in the food industry due to their multifunctional properties. They enhance the stability and bioavailability of nutraceuticals through encapsulation, [302,303] improve texture as texturizing agents, and contribute to the viscosity of formulations. [304–306] Their ability to stabilize emulsions and mimic fat properties supports low-fat product development. [304,305,307–309] Additionally, they serve as structural components, reinforcing mechanical properties [308,310–312] and enhancing barrier properties in biodegradable packaging. [313–315] The development of edible films and 3D scaffolds illustrates their innovative uses. [316–323] Importantly, studies indicate that food-derived amyloid fibrils are safe for human consumption [324] underscoring their potential to revolutionize food product development.

**Table 4** provide a summary of different application of functional amyloid fibrils in Food industry.

**Table 4: application of functional amyloid fibrils in Food industry**

| Application | Description | References |
|---|---|---|
| Encapsulation Agents | Enhance stability and bioavailability of nutraceuticals in food products. | [325,326] |
| Texturizing Agents | Modify rheological properties to improve texture and mouthfeel. | [304,305] |
| Nanoencapsulation of Nutraceuticals | Improve stability and protect bioactive compounds. | [302,303] |
| Thickening Agents | Form gels contributes to viscosity and stability in food formulations. | [303,304,306] |
| Emulsion and Foam Stabilization | Self-assemble nature stabilizes emulsions and foams in food products. | [327–329] |
| Mimicking Fat Properties | Interact with hydrophobic and hydrophilic components to enable fat replacement in low-fat products. | [305,307–309] |
| Structural Components | Provide stability and reinforces mechanical properties in food matrices. | [308,310–312] |
| Biodegradable Packaging | Enhance barrier properties and reduces environmental impact, promoting sustainable packaging solutions. | [313–315] |
| Edible Films and Coatings | Extend shelf life and improves mechanical properties of food products. | [316–322] |
| Cultivated Meat Applications | Create 3D porous scaffolds for cultivated meat applications. | [323] |
| Gastrointestinal Digestion Studies | Have more effective digestion compared to monomers with minimal changes. | [324] |
| Texture Improvement | Improving texture and reducing bitterness by adding fibrils to high-moisture extruded protein isolates | [330] |
| Cold Gelation Procedure | Developed a multistep procedure for $\beta$-lactoglobulin gels that requires lower protein concentrations than conventional methods. | [331] |
| Nanocarriers for Iron Nanoparticles | Act as nanocarriers for iron nanoparticles, providing antioxidant properties and stabilization during digestion. | [332] |
| Enhanced Antioxidant Activity | Improve antioxidant activity, but with different fibrillation potency compared to protein isolate. | [333] |

## 2.14 Amyloid Fibrils as Innovative Solutions for Water and Wastewater Treatment

Amyloid fibrils can be engineered to selectively capture contaminants from water and wastewater due to their ability to form aggregates and bind to various molecules. Tu et al. [334] demonstrated that hierarchical self-assembled amyloid fibrils can reduce membrane fouling in water treatment, enhancing filtration efficiency. Sharma *et al.* [335] showed that these fibrils improve microfiltration membrane performance. Wang *et al.* [336] explored nanoscale amyloid-based methods for water treatment, highlighting their efficacy in dye removal. [337–339] Bai *et al.* [340] discussed covalent organic frameworks for selective dye adsorption, showcasing tailored materials in water treatment. Liu *et al.* [341] developed biodegradable composite beads with amyloid fibrils for wastewater pollutants. Chen *et al.* [342] presented a graphene oxide/amyloid fibrils hydrogel for dye adsorption, while Zhao *et al.* [343] introduced a self-healing hydrogel with amyloid fibrils, indicating promising water treatment applications.

## 2.15 Exploring Amyloid Fibrils for Advanced Energy Storage: From Electrodes to Flexible Devices

Amyloid fibrils, with their unique self-assembling properties and structural stability, are gaining attention for applications in energy storage devices. They serve as promising candidates for electrodes [344–346] due to their high surface area, [12] tunable porosity, [347] and favorable electrochemical properties. [348] Functionalized amyloid fibrils can efficiently store and release energy, leading to enhanced performance in supercapacitors and batteries, including higher energy density, [346] faster charging/discharging rates, [349] and potentially longer cycle life. Their biocompatibility and biodegradability also offer sustainable energy storage solutions. [350] Recent advancements include bio-inspired amyloid composites with high mechanical strength and conductivity for flexible energy storage. [351,352] Lee *et al.* [353] explored self-assembled α-lactalbumin fibrils as supercapacitor electrodes, while Rice *et al.* [354] fabricated amyloid-templated gold microwires for biosensing, demonstrating the versatility of these materials. Despite challenges in scalability and long-term stability, ongoing research is advancing the application of amyloid fibrils in energy storage.

## 2.16 Fibrils as Versatile Scaffolds for Tissue Engineering and Regenerative Medicine

Exploring biocompatible nano-scaffolds for supportive microenvironments significantly enhances cell viability *in-vivo*, advancing cell-based therapies. Various biocompatible materials, including natural polymers, show promise in this area. [355] Peptide and protein self-assembly create biomaterials that mimic the extracellular matrix, promoting cell adhesion, [356–361] migration, [361–365] and differentiation. [361,362,364–366] Zhang *et al.* [367–369] developed self-complementary *β*-sheet peptides that form hydrogels, supporting neuronal cell attachment and differentiation. [357] Self-assembling peptide scaffolds functionalized with osteogenic growth peptide ALK improved proliferation and differentiation of osteoblasts. [360,370] Additionally, amyloid fibrils serve as scaffolds in tissue engineering, [371,372] and enamel repair due to their bioactive surfaces. [373–379] Incorporating natural polyphenols into amyloid fibrils creates hydrogels with unique properties. [380] In tissue engineering, amyloid-based hydrogels provide suitable environments for stem cells, with potential for directing differentiation and enhancing control over stem cell fate [50,381–390]though challenges in tuning stiffness and incorporating nanotopographic cues remain.

## 2.17 Harnessing Amyloid Fibrils for Efficient Light Harvesting

Utilizing amyloid fibrils for light harvesting represents a burgeoning area of research with significant promise. These protein aggregates exhibit intriguing optical properties such as light scattering, [12,391–393] absorption, [394–397] and fluorescence. [244,398–400] Through controllable self-assembly, amyloid fibrils can be engineered to possess desired characteristics, facilitating efficient energy transfer processes crucial [401–404] for light harvesting. Integration with other materials, such as organic dyes [393,405–407] or quantum dots, [408–411] allows for the creation of hybrid structures optimized for various applications, including solar cells and [412–415] LEDs. [416–419] Furthermore, their biocompatibility renders them suitable for biomedical uses. Despite challenges like stability and scalability, ongoing exploration of amyloid fibrils in light harvesting offers exciting prospects for developing novel, high-performance devices with enhanced efficiency and functionality.

## 2.18 Innovative Prospects: Amyloid Fibrils as Catalysts for Nitrogen Catabolism

Amyloid fibrils, traditionally associated with protein misfolding diseases, are now recognized for their potential in nitrogen catabolism, a vital process in biological and environmental systems. [420–422] Researchers explore their ability to mimic or enhance enzyme activity in nitrogen fixation, [356] nitrification [423] and denitrification, [424] leveraging their high surface area [425] and adjustable porosity. [44] By immobilizing enzymes on amyloid fibril scaffolds, they aim to enhance enzyme stability and efficiency. Innovative techniques like directed evolution and protein engineering allow customization for specific applications, promising breakthroughs in agriculture, sustainability, and biotechnology while addressing global challenges like nitrogen pollution and resource management.

## 2.19 Unlocking Cellular Synergy: Exploring Amyloid Fibrils for Enhanced Heterokaryon Compatibility

Amyloid fibrils offer significant potential for advancing biomedical applications, especially in tissue engineering, regenerative medicine, and cell-based therapies. The self-assembling properties of amyloid fibrils are being investigated to facilitate the fusion of different cell types. This strategy could result in the development of hybrid cellular constructs with improved functionality, [44] integrating the advantageous traits of each cell type. These constructs have the potential to offer innovative solutions for tissue regeneration, disease modeling, and drug screening. The inherent biocompatibility and stability of amyloid fibrils also make them promising for enhancing heterokaryon compatibility [426] across different biological systems. While still in its early stages, ongoing research into using amyloid fibrils to promote heterokaryon formation has the potential to enhance our understanding of cell interactions and pave the way for new biomedical innovations.

## 2.20 Unlocking the Potential of Amyloid Fibrils for Advanced Functional Coatings

The use of amyloid fibrils for functional coatings shows great promise across various industries. Functional coatings are essential for improving surface properties like durability, [427] biocompatibility, [428] and functionality, [429] and they have applications that span from biomedical devices to industrial surfaces. Amyloid fibrils, recognized for their unique self-assembling properties and structural stability, present exciting opportunities for developing customized functional coatings. Researchers aim to use the stable and ordered structures of amyloid fibrils to develop coatings with specific functionalities. These coatings could offer antimicrobial properties, improve biocompatibility, or enhance mechanical strength for various surfaces. Amyloid fibrils' biocompatibility and biodegradability make them particularly attractive for biomedical applications. Functional coatings based on amyloid fibrils are crucial for enhancing the performance and compatibility of medical implants, drug delivery systems, and diagnostic devices. As research advances, amyloid fibrils are poised to greatly impact materials science, surface engineering, and biomedical technology. Their use in functional coatings will lead to the development of next-generation coatings with customized properties and improved performance.

## 2.21 Exploring Amyloid Fibrils as Biomimetic Adhesives for Underwater Applications

The use of amyloid fibrils for underwater adhesion [430] is an emerging research area with promising potential across various industries. Although not directly covered in the provided topics, amyloid fibrils offer significant advantages for applications in marine biology, biotechnology, and materials science. Adhesion in water presents challenges due to the disruptive effects of water molecules on traditional adhesives. However, amyloid fibrils' structural stability and adhesive properties provide a strong solution for underwater adhesion. Researchers are exploring how the self-assembling nature of amyloid fibrils can create adhesive materials that form durable bonds with submerged surfaces.

These amyloid-based adhesives could revolutionize fields such as marine engineering, underwater construction, and biomedical implantation. Their biocompatibility and biodegradability further enhance their appeal for biomedical applications, particularly in developing implantable devices and drug delivery systems. As research progresses, amyloid fibrils could open new avenues for advancing technology and tackling challenges in aquatic environments.

## 2.22 Unlocking the Potential of Amyloid Fibrils as Biomimetic Scaffolds for Tissue Engineering

Using amyloid as a scaffold provides unique benefits for promoting cell growth and tissue regeneration. Amyloid fibrils, due to their natural biocompatibility, structural stability, and adjustable properties, offer an excellent platform for creating biomimetic environments that replicate the extracellular matrix (ECM). [44] By harnessing the self-assembling properties of amyloid, researchers can create scaffolds with customized architecture and mechanical properties to direct cell adhesion, [50,431] proliferation, [44,432] and differentiation. [433,434] Additionally, amyloid scaffolds can act as reservoirs for [50,435] and signaling molecules, [436,437] aiding in cellular responses and promoting tissue regeneration. Challenges like degradation kinetics [438,439] and immune response modulation [440,441] must be addressed. However, ongoing research is promising for using amyloid scaffolds to create advanced cell culture systems for tissue engineering. Advancements in amyloid biology and tissue engineering are set to revolutionize regenerative medicine through the use of amyloid scaffolds for cell culture. This approach promises to enable the creation of functional tissues and organs for transplantation and disease modeling.

### 2.23  Exploring Amyloid-Based Materials for Enhanced Efficiency in Organic Photovoltaics

Amyloids show promise for use in organic photovoltaics (OPVs), which are a renewable energy technology. OPVs convert sunlight into electricity, [442] using organic materials and are known for their flexibility, low cost, and lightweight design. Traditionally, OPVs rely on conjugated polymers [443] and small molecules as active materials, but recent research is exploring amyloid-based materials for photovoltaic applications. [444,445]

Amyloids, with their organized nanostructures and semiconducting properties, [381,446] could be effective in OPVs. Researchers are working to harness the self-assembling nature of amyloids to create materials with optimized optoelectronic properties and enhanced charge transport. Additionally, amyloids' biocompatibility and biodegradability offer extra advantages for sustainable energy solutions. Although stability and scalability issues remain, ongoing research is uncovering the potential of amyloids as efficient and eco-friendly materials for OPVs. As our understanding of amyloid biology and photovoltaic technology advances, amyloids could play a significant role in the transition to renewable energy sources.

### 3. Challenges and Future Directions

Addressing the current challenges and limitations in functional amyloid research is crucial for advancing our understanding and fully harnessing the potential of these unique biomaterials. Despite significant progress, a key gap remains in understanding how the structural properties of functional amyloid fibrils relate to their varied biological functions. Resolving these structure-function relationships is essential for designing amyloid-based materials with tailored properties. While functional amyloid fibrils are generally biocompatible, concerns about their potential cytotoxicity, especially in biomedical applications, persist. It is vital to conduct comprehensive biocompatibility assessments and develop strategies to mitigate any adverse effects.

Scaling up production for industrial and clinical applications presents significant challenges. Optimizing processes such as recombinant protein expression and fibril purification is necessary to meet large-scale demands. Precise control over fibril assembly and morphology is also crucial for customizing their properties for specific applications. Developing methods to regulate fibril morphology, size, and surface characteristics will greatly enhance their utility across various fields. The regulatory landscape for amyloid-based materials in biomedical and biotechnological applications is complex. Navigating these regulatory hurdles and securing the necessary approvals are critical for translating research into viable clinical and commercial products.

Addressing these challenges requires interdisciplinary collaborations among researchers in biology, chemistry, materials science, and engineering. Such collaborations can deepen our understanding of functional amyloid fibrils and drive innovation.

Advancements in imaging, spectroscopy, and computational modeling will enable detailed molecular characterization of functional amyloid fibrils, enhancing our understanding of their structure-function dynamics. Bioinspired design strategies will also aid in developing novel amyloid-based materials with improved stability, biocompatibility, and functionality. Additionally, emerging manufacturing technologies—such as microfluidics, 3D printing, and biomimetic assembly—will allow for precise control over fibril assembly and morphology, facilitating scalable production. To transition functional amyloid research from the laboratory to clinical and commercial applications, it is crucial to prioritize rigorous preclinical testing, ensure regulatory compliance, and form strategic partnerships with industry stakeholders. By addressing these challenges and pursuing these future directions, the field of functional amyloid research can unlock new opportunities in biotechnology, medicine, and materials science, leading to transformative advancements with far-reaching impacts.

## 4. Conclusion

While functional and pathological amyloid fibrils share certain similarities, they possess distinct characteristics and fulfill different biological roles. Functional amyloids contribute to various biological processes by forming stable, ordered structures that endure environmental stresses, with their assembly typically regulated and controlled. In contrast, pathological amyloids, linked to protein misfolding diseases, form spontaneously, leading to the buildup of cytotoxic aggregates. Understanding these distinctions is crucial for harnessing the beneficial potential of amyloid fibrils in biotechnological applications beyond disease contexts.

This literature review highlights the extensive applications of functional amyloid fibrils across biotechnology, medicine, materials science, and nanotechnology as shown in **Figure 1**. By thoroughly exploring their structural versatility and biocompatibility, researchers have developed innovative solutions for drug delivery, tissue engineering, and biomaterials design. Additionally, the emerging role of functional amyloids in biological processes underscores their significance as dynamic elements essential for cellular function and organismal survival.

As research continues to advance, the potential for novel applications and groundbreaking discoveries in functional amyloid research remains immense, promising exciting opportunities for transformative impacts on science and technology.

**Figure 1: Industrial application of Functional Amyloid Fibrils**

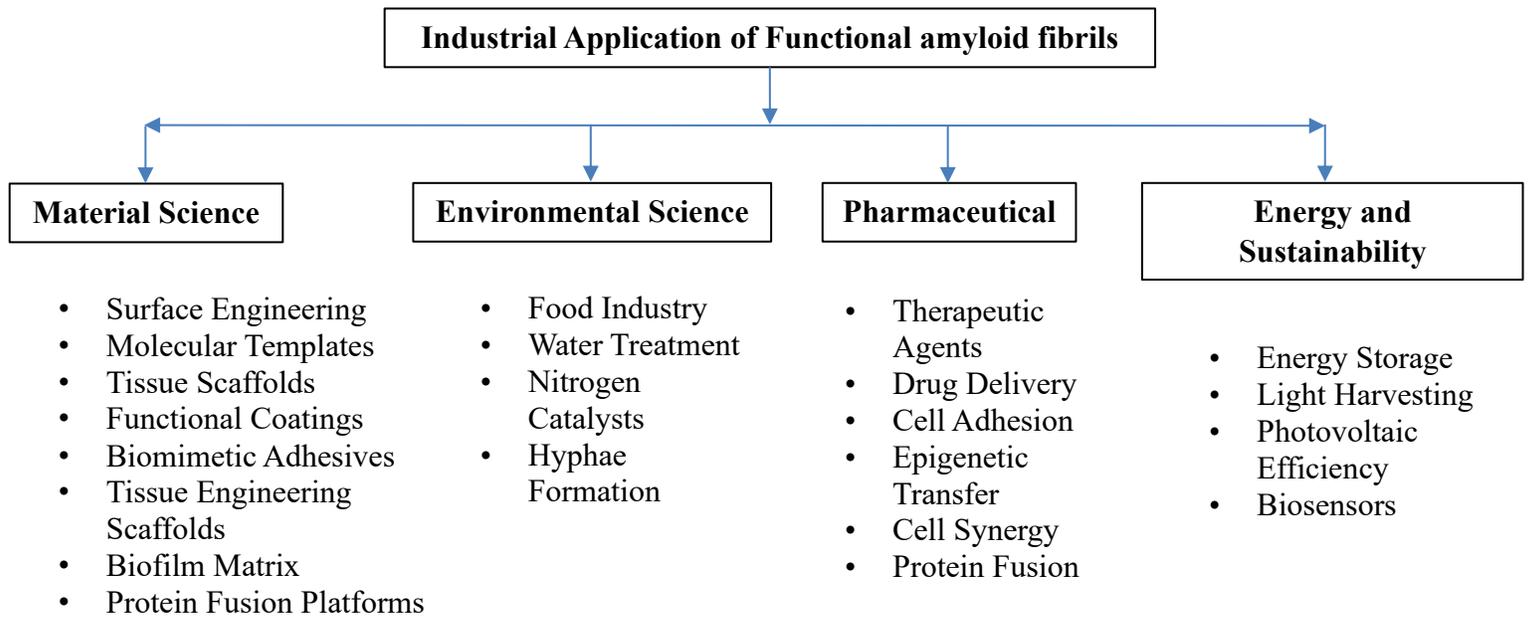